

%
%
%
\def\unredoffs{} \def\redoffs{\voffset=-.31truein\hoffset=-.59truein}
\def\speclscape{\special{ps: landscape}}
%
%
%
%
\newbox\leftpage \newdimen\fullhsize \newdimen\hstitle \newdimen\hsbody
\tolerance=1000\hfuzz=2pt
\catcode`\@=11 
\def\bigans{b }
\def\answ{b }

%
\ifx\answ\bigans\message{(This will come out unreduced.}
\magnification=1200\unredoffs\baselineskip=16pt plus 2pt minus 1pt
\hsbody=\hsize \hstitle=\hsize 
\else\message{(This will be reduced.} \let\l@r=L
\magnification=1000\baselineskip=16pt plus 2pt minus 1pt \vsize=7truein
\redoffs \hstitle=8truein\hsbody=4.75truein\fullhsize=10truein\hsize=\hsbody
\output={\ifnum\pageno=0 
  \shipout\vbox{\speclscape{\hsize\fullhsize\makeheadline}
    \hbox to \fullhsize{\hfill\pagebody\hfill}}\advancepageno
  \else
  \almostshipout{\leftline{\vbox{\pagebody\makefootline}}}\advancepageno
  \fi}
\def\almostshipout#1{\if L\l@r \count1=1 \message{[\the\count0.\the\count1]}
      \global\setbox\leftpage=#1 \global\let\l@r=R
 \else \count1=2
  \shipout\vbox{\speclscape{\hsize\fullhsize\makeheadline}
      \hbox to\fullhsize{\box\leftpage\hfil#1}}  \global\let\l@r=L\fi}
\fi
%
\newcount\yearltd\yearltd=\year\advance\yearltd by -1900

\def\Title#1#2{\nopagenumbers\abstractfont\hsize=\hstitle\rightline{#1}%
\vskip 1in\centerline{\titlefont #2}\abstractfont\vskip .5in\pageno=0}
\def\Date#1{\vfill\leftline{#1}\tenpoint\supereject\global\hsize=\hsbody%
\footline={\hss\tenrm\folio\hss}}
%

\def\draftmode{\message{ DRAFTMODE }\def\draftdate{{\rm preliminary draft:
\number\month/\number\day/\number\yearltd\ \ \hourmin}}%
\headline={\hfil\draftdate}\writelabels\baselineskip=20pt plus 2pt minus 2pt
 {\count255=\time\divide\count255 by 60 \xdef\hourmin{\number\count255}
  \multiply\count255 by-60\advance\count255 by\time
  \xdef\hourmin{\hourmin:\ifnum\count255<10 0\fi\the\count255}}}
\def\nolabels{\def\wrlabeL##1{}\def\eqlabeL##1{}\def\reflabeL##1{}}
\def\writelabels{\def\wrlabeL##1{\leavevmode\vadjust{\rlap{\smash%
{\line{{\escapechar=` \hfill\rlap{\sevenrm\hskip.03in\string##1}}}}}}}%
\def\eqlabeL##1{{\escapechar-1\rlap{\sevenrm\hskip.05in\string##1}}}%
\def\reflabeL##1{\noexpand\llap{\noexpand\sevenrm\string\string\string##1}}}
\nolabels
%
\global\newcount\secno \global\secno=0
\global\newcount\meqno \global\meqno=1
\def\newsec#1{\global\advance\secno by1\message{(\the\secno. #1)}
\global\subsecno=0\eqnres@t\noindent{\bf\the\secno. #1}
\writetoca{{\secsym} {#1}}\par\nobreak\medskip\nobreak}
\def\eqnres@t{\xdef\secsym{\the\secno.}\global\meqno=1\bigbreak\bigskip}
\def\sequentialequations{\def\eqnres@t{\bigbreak}}\xdef\secsym{}
\global\newcount\subsecno \global\subsecno=0
\def\subsec#1{\global\advance\subsecno by1\message{(\secsym\the\subsecno. #1)}
\ifnum\lastpenalty>9000\else\bigbreak\fi
\noindent{\it\secsym\the\subsecno. #1}\writetoca{\string\quad
{\secsym\the\subsecno.} {#1}}\par\nobreak\medskip\nobreak}
\def\appendix#1#2{\global\meqno=1\global\subsecno=0\xdef\secsym{\hbox{#1.}}
\bigbreak\bigskip\noindent{\bf Appendix #1. #2}\message{(#1. #2)}
\writetoca{Appendix {#1.} {#2}}\par\nobreak\medskip\nobreak}
%
%
\def\eqnn#1{\xdef #1{(\secsym\the\meqno)}\writedef{#1\leftbracket#1}%
\global\advance\meqno by1\wrlabeL#1}
\def\eqna#1{\xdef #1##1{\hbox{$(\secsym\the\meqno##1)$}}
\writedef{#1\numbersign1\leftbracket#1{\numbersign1}}%
\global\advance\meqno by1\wrlabeL{#1$\{\}$}}
\def\eqn#1#2{\xdef #1{(\secsym\the\meqno)}\writedef{#1\leftbracket#1}%
\global\advance\meqno by1$$#2\eqno#1\eqlabeL#1$$}
%
\newskip\footskip\footskip14pt plus 1pt minus 1pt 
\def\footnotefont{\ninepoint}\def\f@t#1{\footnotefont #1\@foot}
\def\f@@t{\baselineskip\footskip\bgroup\footnotefont\aftergroup\@foot\let\next}
\setbox\strutbox=\hbox{\vrule height9.5pt depth4.5pt width0pt}
\global\newcount\ftno \global\ftno=0
\def\foot{\global\advance\ftno by1\footnote{$^{\the\ftno}$}}
%
\newwrite\ftfile
\def\footend{\def\foot{\global\advance\ftno by1\chardef\wfile=\ftfile
$^{\the\ftno}$\ifnum\ftno=1\immediate\openout\ftfile=foots.tmp\fi%
\immediate\write\ftfile{\noexpand\smallskip%
\noexpand\item{f\the\ftno:\ }\pctsign}\findarg}%
\def\footatend{\vfill\eject\immediate\closeout\ftfile{\parindent=20pt
\centerline{\bf Footnotes}\nobreak\bigskip\input foots.tmp }}}
\def\footatend{}
%
%
\global\newcount\refno \global\refno=1
\newwrite\rfile
\def\ref{[\the\refno]\nref}
\def\nref#1{\xdef#1{[\the\refno]}\writedef{#1\leftbracket#1}%
\ifnum\refno=1\immediate\openout\rfile=refs.tmp\fi
\global\advance\refno by1\chardef\wfile=\rfile\immediate
\write\rfile{\noexpand\item{#1\ }\reflabeL{#1\hskip.31in}\pctsign}\findarg}
\def\findarg#1#{\begingroup\obeylines\newlinechar=`\^^M\pass@rg}
{\obeylines\gdef\pass@rg#1{\writ@line\relax #1^^M\hbox{}^^M}%
\gdef\writ@line#1^^M{\expandafter\toks0\expandafter{\striprel@x #1}%
\edef\next{\the\toks0}\ifx\next\em@rk\let\next=\endgroup\else\ifx\next\empty%
\else\immediate\write\wfile{\the\toks0}\fi\let\next=\writ@line\fi\next\relax}}
\def\striprel@x#1{} \def\em@rk{\hbox{}}
\def\lref{\begingroup\obeylines\lr@f}
\def\lr@f#1#2{\gdef#1{\ref#1{#2}}\endgroup\unskip}

\def\addref#1{\immediate\write\rfile{\noexpand\item{}#1}} 
\def\startrefs#1{\immediate\openout\rfile=refs.tmp\refno=#1}
\def\xref{\expandafter\xr@f}\def\xr@f[#1]{#1}
\def\refs#1{\count255=1[\r@fs #1{\hbox{}}]}
\def\r@fs#1{\ifx\und@fined#1\message{reflabel \string#1 is undefined.}%
\nref#1{need to supply reference \string#1.}\fi%
\vphantom{\hphantom{#1}}\edef\next{#1}\ifx\next\em@rk\def\next{}%
\else\ifx\next#1\ifodd\count255\relax\xref#1\count255=0\fi%
\else#1\count255=1\fi\let\next=\r@fs\fi\next}
%

%
\newwrite\ffile\global\newcount\figno \global\figno=1
\def\fig{fig.~\the\figno\nfig}
\def\nfig#1{\xdef#1{fig.~\the\figno}%
\writedef{#1\leftbracket fig.\noexpand~\the\figno}%
\ifnum\figno=1\immediate\openout\ffile=figs.tmp\fi\chardef\wfile=\ffile%
\immediate\write\ffile{\noexpand\medskip\noexpand\item{Fig.\ \the\figno. }
\reflabeL{#1\hskip.55in}\pctsign}\global\advance\figno by1\findarg}
\def\xfig{\expandafter\xf@g}\def\xf@g fig.\penalty\@M\ {}
\def\figs#1{figs.~\f@gs #1{\hbox{}}}
\def\f@gs#1{\edef\next{#1}\ifx\next\em@rk\def\next{}\else
\ifx\next#1\xfig #1\else#1\fi\let\next=\f@gs\fi\next}
\newwrite\lfile
{\escapechar-1\xdef\pctsign{\string\%}\xdef\leftbracket{\string\{}
\xdef\rightbracket{\string\}}\xdef\numbersign{\string\#}}

\def\writestop{\def\writestoppt{\immediate\write\lfile{\string\pageno%
\the\pageno\string\startrefs\leftbracket\the\refno\rightbracket%
\string\def\string\secsym\leftbracket\secsym\rightbracket%
\string\secno\the\secno\string\meqno\the\meqno}\immediate\closeout\lfile}}
\def\writestoppt{}\def\writedef#1{}
\def\seclab#1{\xdef #1{\the\secno}\writedef{#1\leftbracket#1}\wrlabeL{#1=#1}}
\def\subseclab#1{\xdef #1{\secsym\the\subsecno}%
\writedef{#1\leftbracket#1}\wrlabeL{#1=#1}}
\newwrite\tfile \def\writetoca#1{}
\def\leaderfill{\leaders\hbox to 1em{\hss.\hss}\hfill}
\def\writetoc{\immediate\openout\tfile=toc.tmp
   \def\writetoca##1{{\edef\next{\write\tfile{\noindent ##1
   \string\leaderfill {\noexpand\number\pageno} \par}}\next}}}

%
\catcode`\@=12 
%
\edef\tfontsize{\ifx\answ\bigans scaled\magstep3\else scaled\magstep4\fi}
\font\titlerm=cmr10 \tfontsize \font\titlerms=cmr7 \tfontsize
\font\titlermss=cmr5 \tfontsize \font\titlei=cmmi10 \tfontsize
\font\titleis=cmmi7 \tfontsize \font\titleiss=cmmi5 \tfontsize
\font\titlesy=cmsy10 \tfontsize \font\titlesys=cmsy7 \tfontsize
\font\titlesyss=cmsy5 \tfontsize \font\titleit=cmti10 \tfontsize
\skewchar\titlei='177 \skewchar\titleis='177 \skewchar\titleiss='177
\skewchar\titlesy='60 \skewchar\titlesys='60 \skewchar\titlesyss='60
\def\titlefont{\def\rm{\fam0\titlerm}
\textfont0=\titlerm \scriptfont0=\titlerms \scriptscriptfont0=\titlermss
\textfont1=\titlei \scriptfont1=\titleis \scriptscriptfont1=\titleiss
\textfont2=\titlesy \scriptfont2=\titlesys \scriptscriptfont2=\titlesyss
\textfont\itfam=\titleit \def\it{\fam\itfam\titleit}\rm}
 \ifx\answ\bigans\else scaled\magstep1\fi
\ifx\answ\bigans\def\abstractfont{\tenpoint}\else
\font\abssl=cmsl10 scaled \magstep1
\font\absrm=cmr10 scaled\magstep1 \font\absrms=cmr7 scaled\magstep1
\font\absrmss=cmr5 scaled\magstep1 \font\absi=cmmi10 scaled\magstep1
\font\absis=cmmi7 scaled\magstep1 \font\absiss=cmmi5 scaled\magstep1
\font\abssy=cmsy10 scaled\magstep1 \font\abssys=cmsy7 scaled\magstep1
\font\abssyss=cmsy5 scaled\magstep1 \font\absbf=cmbx10 scaled\magstep1
\skewchar\absi='177 \skewchar\absis='177 \skewchar\absiss='177
\skewchar\abssy='60 \skewchar\abssys='60 \skewchar\abssyss='60
\def\abstractfont{\def\rm{\fam0\absrm}
\textfont0=\absrm \scriptfont0=\absrms \scriptscriptfont0=\absrmss
\textfont1=\absi \scriptfont1=\absis \scriptscriptfont1=\absiss
\textfont2=\abssy \scriptfont2=\abssys \scriptscriptfont2=\abssyss
\textfont\itfam=\bigit \def\it{\fam\itfam\bigit}\def\footnotefont{\tenpoint}%
\textfont\slfam=\abssl \def\sl{\fam\slfam\abssl}%
\textfont\bffam=\absbf \def\bf{\fam\bffam\absbf}\rm}\fi
\def\tenpoint{\def\rm{\fam0\tenrm}
\textfont0=\tenrm \scriptfont0=\sevenrm \scriptscriptfont0=\fiverm
\textfont1=\teni  \scriptfont1=\seveni  \scriptscriptfont1=\fivei
\textfont2=\tensy \scriptfont2=\sevensy \scriptscriptfont2=\fivesy
\textfont\itfam=\tenit \def\it{\fam\itfam\tenit}\def\footnotefont{\ninepoint}%
\textfont\bffam=\tenbf \def\bf{\fam\bffam\tenbf}\def\sl{\fam\slfam\tensl}\rm}
\font\ninerm=cmr9 \font\sixrm=cmr6 \font\ninei=cmmi9 \font\sixi=cmmi6
\font\ninesy=cmsy9 \font\sixsy=cmsy6 \font\ninebf=cmbx9
\font\nineit=cmti9 \font\ninesl=cmsl9 \skewchar\ninei='177
\skewchar\sixi='177 \skewchar\ninesy='60 \skewchar\sixsy='60
\def\ninepoint{\def\rm{\fam0\ninerm}
\textfont0=\ninerm \scriptfont0=\sixrm \scriptscriptfont0=\fiverm
\textfont1=\ninei \scriptfont1=\sixi \scriptscriptfont1=\fivei
\textfont2=\ninesy \scriptfont2=\sixsy \scriptscriptfont2=\fivesy
\textfont\itfam=\ninei \def\it{\fam\itfam\nineit}\def\sl{\fam\slfam\ninesl}%
\textfont\bffam=\ninebf \def\bf{\fam\bffam\ninebf}\rm}
%
%

\hyphenation{anom-aly anom-alies coun-ter-term coun-ter-terms}
\def\inv{^{\raise.15ex\hbox{${\scriptscriptstyle -}$}\kern-.05em 1}}

\def\Dsl{\,\raise.15ex\hbox{/}\mkern-13.5mu D} 
\def\dsl{\raise.15ex\hbox{/}\kern-.57em\partial}

\font\bigit=cmti10 scaled \magstep1
\def\lspace{\ifx\answ\bigans{}\else\qquad\fi}
\def\lbspace{\ifx\answ\bigans{}\else\hskip-.2in\fi} 
\def\boxeqn#1{\vcenter{\vbox{\hrule\hbox{\vrule\kern3pt\vbox{\kern3pt
	\hbox{${\displaystyle #1}$}\kern3pt}\kern3pt\vrule}\hrule}}}
\def\mbox#1#2{\vcenter{\hrule \hbox{\vrule height#2in
		\kern#1in \vrule} \hrule}}  
%
 \def\CC{{\cal C}}

\def\vev#1{\langle #1 \rangle}

\def\darr#1{\raise1.5ex\hbox{$\leftrightarrow$}\mkern-16.5mu #1}

\def\roughly#1{\raise.3ex\hbox{$#1$\kern-.75em\lower1ex\hbox{$\sim$}}}

\let\includefigures=\iftrue
\let\useblackboard=\iftrue
\newfam\black

\includefigures
\message{If you do not have epsf.tex (to include figures),}
\message{change the option at the top of the tex file.}
\input epsf
\def\figin{\epsfcheck\figin}\def\figins{\epsfcheck\figins}
\def\epsfcheck{\ifx\epsfbox\UnDeFiNeD
\message{(NO epsf.tex, FIGURES WILL BE IGNORED)}
\gdef\figin##1{\vskip2in}\gdef\figins##1{\hskip.5in}
\else\message{(FIGURES WILL BE INCLUDED)}%
\gdef\figin##1{##1}\gdef\figins##1{##1}\fi}
\def\DefWarn#1{}
\def\figinsert{\goodbreak\midinsert}
\def\ifig#1#2#3{\DefWarn#1\xdef#1{fig.~\the\figno}
\writedef{#1\leftbracket fig.\noexpand~\the\figno}%
\figinsert\figin{\centerline{#3}}\medskip\centerline{\vbox{
\baselineskip12pt\advance\hsize by -1truein
\noindent\footnotefont{\bf Fig.~\the\figno:} #2}}
\endinsert\global\advance\figno by1}
\else
\def\ifig#1#2#3{\xdef#1{fig.~\the\figno}
\writedef{#1\leftbracket fig.\noexpand~\the\figno}%
\global\advance\figno by1} \fi

\def\id{{1 \kern-.28em {\rm l}}}

\def\K3{{\bf K3}}
\def\journal#1&#2(#3){\unskip, \sl #1\ \bf #2 \rm(19#3) }
\def\andjournal#1&#2(#3){\sl #1~\bf #2 \rm (19#3) }

\def\hat{\widehat}
\def\ie{{\it i.e.}}

\def\tilde{\widetilde}

\def\frac#1#2{{#1\over#2}}

\def\ket#1{|#1\rangle}
\def\bra#1{\langle#1|}
\def\vev#1{\langle#1\rangle}

\def\inbar{\,\vrule height1.5ex width.4pt depth0pt}
\def\IC{\relax\hbox{$\inbar\kern-.3em{\rm C}$}}
\def\IR{\relax{\rm I\kern-.18em R}}
\def\IP{\relax{\rm I\kern-.18em P}}

%
%

%
\catcode`\@=11
\def\slash#1{\mathord{\mathpalette\c@ncel{#1}}}
\overfullrule=0pt
\def\AA{{\cal A}}
\def\BB{{\cal B}}
\def\CC{{\cal C}}

\def\LL{{\cal L}}
\def\NN{{\cal N}}
\def\OO{{\cal O}}

\def\underrel#1\over#2{\mathrel{\mathop{\kern\z@#1}\limits_{#2}}}

\catcode`\@=12


%
\def\bra#1{\left\langle #1\right|}
\def\ket#1{\left| #1\right\rangle}
\def\vev#1{\left\langle #1 \right\rangle}
\def\det{{\rm det}}

\def\det{{\rm det}}


\def\p{{\partial}}

\def\ra{{\rightarrow}}

\def\tq{{\tilde{q}}}
\def\tom{{\tilde{\omega}}}

\def\LL{{\cal L}}


\lref\MetsaevTseytlina{
  R.~R.~Metsaev and A.~A.~Tseytlin,
  ``Curvature Cubed Terms in String Theory Effective Actions,''
  Phys.\ Lett.\  B {\bf 185}, 52 (1987).
}

\lref\BastianelliFTa{
  F.~Bastianelli, S.~Frolov and A.~A.~Tseytlin,
  ``Three-point correlators of stress tensors in maximally-supersymmetric
  conformal theories in d = 3 and d = 6,''
  Nucl.\ Phys.\  B {\bf 578}, 139 (2000)
  [arXiv:hep-th/9911135].
}

\lref\BastianelliFTb{
  F.~Bastianelli, S.~Frolov and A.~A.~Tseytlin,
  ``Conformal anomaly of (2,0) tensor multiplet in six dimensions and  AdS/CFT
  correspondence,''
  JHEP {\bf 0002}, 013 (2000)
  [arXiv:hep-th/0001041].
}


\lref\BuchelMyers{
  A.~Buchel and R.~C.~Myers,
  ``Causality of Holographic Hydrodynamics,''
  arXiv:0906.2922 [hep-th].
}

\lref\BriganteLMSYa{
  M.~Brigante, H.~Liu, R.~C.~Myers, S.~Shenker and S.~Yaida,
  ``Viscosity Bound Violation in Higher Derivative Gravity,''
  Phys.\ Rev.\  D {\bf 77}, 126006 (2008)
  [arXiv:0712.0805 [hep-th]].
}

\lref\BriganteLMSYb{
  M.~Brigante, H.~Liu, R.~C.~Myers, S.~Shenker and S.~Yaida,
  ``The Viscosity Bound and Causality Violation,''
  Phys.\ Rev.\ Lett.\  {\bf 100}, 191601 (2008)
  [arXiv:0802.3318 [hep-th]].
}


\lref\Hofman{
  D.~M.~Hofman,
  ``Higher Derivative Gravity, Causality and Positivity of Energy in a UV
  complete QFT,''
  arXiv:0907.1625 [hep-th].
}

\lref\HofmanMaldacena{
  D.~M.~Hofman and J.~Maldacena,
  ``Conformal collider physics: Energy and charge correlations,''
  JHEP {\bf 0805}, 012 (2008)
  [arXiv:0803.1467 [hep-th]].
}


\lref\DehghaniPourhasan{
  M.~H.~Dehghani and R.~Pourhasan,
  ``Thermodynamic instability of black holes of third order Lovelock gravity,''
  Phys.\ Rev.\  D {\bf 79}, 064015 (2009)
  [arXiv:0903.4260 [gr-qc]].
}

\lref\DehghaniBH{
  M.~H.~Dehghani, N.~Bostani and S.~H.~Hendi,
  ``Magnetic Branes in Third Order Lovelock-Born-Infeld Gravity,''
  Phys.\ Rev.\  D {\bf 78}, 064031 (2008)
  [arXiv:0806.1429 [gr-qc]].
}

\lref\DehghaniAH{
  M.~H.~Dehghani, N.~Alinejadi and S.~H.~Hendi,
  ``Topological Black Holes in Lovelock-Born-Infeld Gravity,''
  Phys.\ Rev.\  D {\bf 77}, 104025 (2008)
  [arXiv:0802.2637 [hep-th]].
}

\lref\Lovelock{
  D.~Lovelock
  ``The Einstein Tensor and Its Generalizations''
  J.Math.Phys. \ {\bf 12}, 498 (1971)
}

\lref\NojiriMH{
  S.~Nojiri and S.~D.~Odintsov,
  ``On the conformal anomaly from higher derivative gravity in AdS/CFT
  correspondence,''
  Int.\ J.\ Mod.\ Phys.\  A {\bf 15}, 413 (2000)
  [arXiv:hep-th/9903033].
}

\lref\HS{
  M.~Henningson and K.~Skenderis,
  ``The holographic Weyl anomaly,''
  JHEP {\bf 9807}, 023 (1998)
  [arXiv:hep-th/9806087];
  ``Holography and the Weyl anomaly,''
  Fortsch.\ Phys.\  {\bf 48}, 125 (2000)
  [arXiv:hep-th/9812032].
}

\lref\OsbornErdmenger{
  J.~Erdmenger and H.~Osborn,
  ``Conserved currents and the energy-momentum tensor in conformally  invariant
  theories for general dimensions,''
  Nucl.\ Phys.\  B {\bf 483}, 431 (1997)
  [arXiv:hep-th/9605009].
}

\lref\OsbornPetkou{
  H.~Osborn and A.~C.~Petkou,
  ``Implications of Conformal Invariance in Field Theories for General
  Dimensions,''
  Annals Phys.\  {\bf 231}, 311 (1994)
  [arXiv:hep-th/9307010].
}

\lref\ParnachevYT{
  A.~Parnachev and S.~S.~Razamat,
  ``Comments on Bounds on Central Charges in N=1 Superconformal Theories,''
  JHEP {\bf 0907}, 010 (2009)
  [arXiv:0812.0781 [hep-th]].
}

\lref\BanerjeeFM{
  N.~Banerjee and S.~Dutta,
  JHEP {\bf 0907}, 024 (2009)
  [arXiv:0903.3925 [hep-th]].
}

\lref\BuchelVZ{
  A.~Buchel, R.~C.~Myers and A.~Sinha,
  JHEP {\bf 0903}, 084 (2009)
  [arXiv:0812.2521 [hep-th]].
}

\lref\AdamsZK{
  A.~Adams, A.~Maloney, A.~Sinha and S.~E.~Vazquez,
  ``1/N Effects in Non-Relativistic Gauge-Gravity Duality,''
  JHEP {\bf 0903}, 097 (2009)
  [arXiv:0812.0166 [hep-th]].
}

\lref\ShuAX{
  F.~W.~Shu,
  ``The Quantum Viscosity Bound In Lovelock Gravity,''
  arXiv:0910.0607 [hep-th].
}

\lref\GeAC{
  X.~H.~Ge, S.~J.~Sin, S.~F.~Wu and G.~H.~Yang,
  ``Shear viscosity and instability from third order Lovelock gravity,''
  arXiv:0905.2675 [hep-th].
}

\lref\BanerjeeFM{
  N.~Banerjee and S.~Dutta,
  ``Shear Viscosity to Entropy Density Ratio in Six Derivative Gravity,''
  JHEP {\bf 0907}, 024 (2009)
  [arXiv:0903.3925 [hep-th]].
}

\lref\CremoniniSY{
  S.~Cremonini, K.~Hanaki, J.~T.~Liu and P.~Szepietowski,
  ``Higher derivative effects on eta/s at finite chemical potential,''
  Phys.\ Rev.\  D {\bf 80}, 025002 (2009)
  [arXiv:0903.3244 [hep-th]].
}

\lref\GeEH{
  X.~H.~Ge and S.~J.~Sin,
  ``Shear viscosity, instability and the upper bound of the Gauss-Bonnet
  coupling constant,''
  JHEP {\bf 0905}, 051 (2009)
  [arXiv:0903.2527 [hep-th]].
}

\lref\BrusteinCG{
  R.~Brustein and A.~J.~M.~Medved,
  ``The ratio of shear viscosity to entropy density in generalized theories of
  gravity,''
  Phys.\ Rev.\  D {\bf 79}, 021901 (2009)
  [arXiv:0808.3498 [hep-th]].
}

\lref\KatsMQ{
  Y.~Kats and P.~Petrov,
  ``Effect of curvature squared corrections in AdS on the viscosity of the dual
  gauge theory,''
  JHEP {\bf 0901}, 044 (2009)
  [arXiv:0712.0743 [hep-th]].
}

\lref\KSS{
  P.~Kovtun, D.~T.~Son and A.~O.~Starinets,
  ``Viscosity in strongly interacting quantum field theories from black hole
  Phys.\ Rev.\ Lett.\  {\bf 94}, 111601 (2005)
  [arXiv:hep-th/0405231].
  ``Holography and hydrodynamics: Diffusion on stretched horizons,''
  JHEP {\bf 0310}, 064 (2003)
  [arXiv:hep-th/0309213].
}

\lref\IntriligatorJJ{
  K.~A.~Intriligator and B.~Wecht,
  ``The exact superconformal R-symmetry maximizes a,''
  Nucl.\ Phys.\  B {\bf 667}, 183 (2003)
  [arXiv:hep-th/0304128].
}

\lref\KutasovIY{
  D.~Kutasov, A.~Parnachev and D.~A.~Sahakyan,
  ``Central charges and U(1)R symmetries in N = 1 super Yang-Mills,''
  JHEP {\bf 0311}, 013 (2003)
  [arXiv:hep-th/0308071].
}

\lref\NeupaneDC{
  I.~P.~Neupane and N.~Dadhich,
  ``Higher Curvature Gravity: Entropy Bound and Causality Violation,''
  arXiv:0808.1919 [hep-th].
}

\lref\BuchelVZ{
  A.~Buchel, R.~C.~Myers and A.~Sinha,
  ``Beyond eta/s = 1/4pi,''
  JHEP {\bf 0903}, 084 (2009)
  [arXiv:0812.2521 [hep-th]].
}

\lref\PalQG{
  S.~S.~Pal,
  ``$\eta/s$ at finite coupling,''
  arXiv:0910.0101 [hep-th].
}

\lref\WP{
Work in progress.  }

\lref\SinhaMyers{
   A.~Sinha and R.~Myers
  ``The viscosity bound in string theory,''
   arXiv:0907.4798 [hep-th]
}

\lref\SveshikovTkachov{
  N.~A.~Sveshnikov and F.~V.~Tkachov,
  ``Jets and quantum field theory,''
  Phys.\ Lett.\  B {\bf 382} (1996) 403
  [arXiv:hep-ph/9512370].
}

\lref\ShapereTachikawa{
  A.~D.~Shapere and Y.~Tachikawa,
  ``Central charges of N=2 superconformal field theories in four dimensions,''
  JHEP {\bf 0809}, 109 (2008)
  [arXiv:0804.1957 [hep-th]].
}

\lref\BoulwareDeser{
  D.~G.~Boulware and S.~Deser,
  ``String Generated Gravity Models,''
  Phys.\ Rev.\ Lett.\  {\bf 55}, 2656 (1985).
}

\lref\Myers{
  R.~C.~Myers,
  ``HIGHER DERIVATIVE GRAVITY, SURFACE TERMS AND STRING THEORY,''
  Phys.\ Rev.\  D {\bf 36}, 392 (1987).
}

\lref\Zumino{
  B.~Zumino,
  ``Gravity Theories In More Than Four-Dimensions,''
  Phys.\ Rept.\  {\bf 137}, 109 (1986).
}

\lref\Cai{
  R.~G.~Cai,
  ``Gauss-Bonnet black holes in AdS spaces,''
  Phys.\ Rev.\  D {\bf 65}, 084014 (2002)
  [arXiv:hep-th/0109133].
}

\lref\ExirifardJabbari{
  Q.~Exirifard and M.~M.~Sheikh-Jabbari,
  ``Lovelock Gravity at the Crossroads of Palatini and Metric Formulations,''
  Phys.\ Lett.\  B {\bf 661}, 158 (2008)
  [arXiv:0705.1879 [hep-th]].
}

\lref\Cremoninia{
  S.~Cremonini, J.~T.~Liu and P.~Szepietowski,
  ``Higher Derivative Corrections to R-charged Black Holes: Boundary
  Counterterms and the Mass-Charge Relation,''
  arXiv:0910.5159 [hep-th].
}

\lref\Paulos{
  M.~F.~Paulos,
  ``Transport coefficients, membrane couplings and universality at
 extremality,''
  arXiv:0910.4602 [hep-th].
}

\lref\Caib{
  R.~G.~Cai, Y.~Liu and Y.~W.~Sun,
  ``Transport Coefficients from Extremal Gauss-Bonnet Black Holes,''
  arXiv:0910.4705 [hep-th].
}

\lref\Ohta{R.~G.~Cai, Z.~Y.~Nie, N.~Ohta and Y.~W.~Sun,
  ``Shear Viscosity from Gauss-Bonnet Gravity with a Dilaton Coupling,''
  Phys.\ Rev.\  D {\bf 79} (2009) 066004
  [arXiv:0901.1421 [hep-th]].
}

\lref\OsbornQU{
  H.~Osborn,
  ``N = 1 superconformal symmetry in four-dimensional quantum field theory,''
  Annals Phys.\  {\bf 272}, 243 (1999)
  [arXiv:hep-th/9808041].
}

\Title{\vbox{\baselineskip12pt
}}
{\vbox{\centerline{$AdS_7/CFT_6$, Gauss-Bonnet Gravity, and Viscosity Bound}
\vskip.06in
}}
\centerline{Jan de Boer${}^a$, Manuela Kulaxizi${}^a$, and Andrei Parnachev${}^b$}
\bigskip
\centerline{{\it ${}^a$Department of Physics, University of Amsterdam }}
\centerline{{\it Valckenierstraat 65, 1018XE Amsterdam, The Netherlands }}
\centerline{{\it ${}^b$C.N.Yang Institute for Theoretical Physics, Stony Brook University}}
\centerline{{\it Stony Brook, NY 11794-3840, USA}}
\vskip.1in \vskip.1in \centerline{\bf Abstract}
\noindent
We study the relation between the causality and positivity of energy bounds for
Gauss-Bonnet gravity in an $AdS_7$ background.
Requiring the group velocity of metastable states to be bounded by the speed of light
places a bound on the value of Gauss-Bonnet coupling.
To find the positivity of energy constraints we compute the parameters
which determine the angular distribution of the energy flux in terms
of three independent coefficients specifying the three-point function of the
stress-energy tensor.
We then relate the latter to the Weyl anomaly of the six-dimensional CFT
and compute the anomaly holographically.
The resulting upper bound on the Gauss-Bonnet coupling coincides with that from
causality and results in a new bound on viscosity/entropy ratio.
\vfill

\Date{January 2010}


\newsec{Introduction and summary}

\noindent Recently an intriguing connection between
positivity of energy flux and causality has been addressed
in the context of higher derivative gravity \refs{\HofmanMaldacena\BriganteLMSYa\BriganteLMSYb\BuchelMyers-\Hofman}.
Consider Gauss-Bonnet gravity with negative cosmological constant
in five dimensions.
Since AdS space is a solution of the equations of motion, one can
hypothesize the existence of a dual CFT.
The theory furthermore possesses exact black hole solutions which are
asymptotically AdS.
Brigante, Liu, Myers, Shenker and Yaida \refs{\BriganteLMSYa-\BriganteLMSYb}
considered gravitons propagating in these backgrounds and
found long lived excitations which correspond to metastable states in the boundary theory.
Restricting the group velocity of these states to be bounded by the speed
of light (requiring causality of the boundary theory at finite temperature) places
non-trivial constraints on the value of Gauss-Bonnet coupling $\lambda$ \refs{\BriganteLMSYa-\BriganteLMSYb}.

At first sight, a completely unrelated set of constraints was proposed
for CFTs by Hofman and Maldacena \HofmanMaldacena.
By requiring positivity of the energy flux measured in a collision of certain
CFT states, they deduced a set of constraints on
the quantities $t_2$ and $t_4$ which determine the angular distribution of the
energy flux.
The values of $t_2$ and $t_4$ are completely determined
by the two- and three-point functions of the stress energy tensor.
The constraints  can be reformulated as
bounds on the ratio $a^{(d=4)}/c^{(d=4)}$ of the coefficients which appear in the Weyl anomaly of a
four-dimensional CFT.
In the supersymmetric case $t_4=0$ and the bounds are
\eqn\susybounds{  {1\over2}\leq {a^{(d=4)}\over c^{(d=4)}}  \leq {3\over2}   }
These bounds have been verified for a number of interacting
superconformal theories \ParnachevYT.
(More stringent bounds exist for ${\cal N}=2$ superconformal four dimensional field theories
 \ShapereTachikawa .)
In the non-supersymmetric case $a^{(d=4)}$ and $c^{(d=4)}$ do not completely determine
$t_2$ and $t_4$, and therefore the bounds on $a^{(d=4)}/c^{(d=4)}$ are slightly relaxed.
In \HofmanMaldacena\ it has been noted that for Gauss-Bonnet gravity
the lower bound in  \susybounds\ precisely translates into the upper bound on $\lambda$
coming from causality\foot{The upper bound on $\lambda$ implies a new bound
on the viscosity/entropy ratio in higher derivative gravity \refs{\BriganteLMSYa-\BriganteLMSYb}.
The original KSS bound $\eta/s\geq 1/4\pi$ \KSS\ has been shown to be violated
in a controlled setting \KatsMQ. Other recent work on the higher derivative
corrections to $\eta/s$ includes \refs{\NeupaneDC\BrusteinCG\AdamsZK\BuchelVZ\GeEH\CremoniniSY\BanerjeeFM\GeAC\PalQG\ShuAX\Paulos\Caib\Ohta-\Cremoninia}.}.
Recently the upper bound in \susybounds\ has been reproduced by examining
a differently polarized graviton (shear channel) \BuchelMyers
(see also \Hofman).

At first sight this relation is  puzzling.
Why would the $\NN=1$ superconformal result have anything to
do with Gauss-Bonnet gravity?
It is worth to note however that the $R^3$ terms are absent in
supersymmetric string theories, so one may fantasize that
some supersymmetric string compactification would yield Gauss-Bonnet
gravity as the low energy theory.
In this paper we investigate the state of the correspondence between
causality and positivity of the energy flux
in the case of Gauss-Bonnet (GB) gravity in an $AdS_7$ background possibly
dual to some six-dimensional CFT.
An important feature of the six-dimensional CFT is that,
unlike the four-dimensional case, the knowledge of the Weyl anomaly
completely determines the values of $t_2$ and $t_4$.
In particular, we find that $t_4=0$.
Note that vanishing $t_4$ is a necessary feature of  any supersymmetric CFT.
We compute the upper bound on $\lambda$ by requiring both causality
of the boundary theory at finite temperature and positivity of the energy flux.
The two bounds coincide exactly and result in a new bound on $\eta/s$.
We also compute the lower bound on $\lambda$.

The rest of the paper is organized as follows.
In the next Section we review Gauss-Bonnet gravity and asymptotically
AdS black holes solutions.
In Section 3 we study small fluctuations around these black holes,
and find metastable states which can lead to causality violation.
By requiring the group velocity of these states to be bounded by the speed
of light, we find an upper bound on the value of $\lambda$.
In Section 4 we generalize the results of \HofmanMaldacena\ to six
dimensions.
We compute the values of $t_2$ and $t_4$ in terms of the parameters
$a,b,c$ which determine the two- and three--point functions of the stress-energy tensor.
In Section 5 we compute the coefficients $a,b,c$ (and hence $t_2$ and $t_4$)
in GB gravity.
We do it by relating the values of $a,b,c$ to the coefficients of the
B-type terms in the Weyl anomaly and computing the anomaly holographically.
We then find the bounds on $\lambda$ and observe that the upper
bound is precisely the same as the one found in Section 3.
We discuss our results in Section 6.
Some technical results appear in the Appendix.

\newsec{Gauss-Bonnet gravity}

\noindent
Among gravity theories which involve higher derivative terms of
the Riemann tensor in their actions, there is a special
class which shares many of the properties of Einstein-Hilbert gravity.
It is usually referred to as Lovelock gravity \refs{\Lovelock\Zumino-\Myers} and is the
most general theory of gravitation whose equations of motion
contain at most second order derivatives of the metric.
Recently, the Palatini and metric formulations of Lovelock gravity
have been shown to be equivalent \ExirifardJabbari.

The Lovelock action for a $d+1$-dimensional spacetime can be written as
\eqn\LovelockAction{S={1\over l_P^{d-1}} \int d^{{d+1}}x \sqrt{-g}\sum_{p=1}^{[{d\over 2}]} \lambda_{p} {\cal{L}}_{p}}
Here $l_P$ is Planck's length, $[{d\over 2}]$ denotes the integral part of ${d\over 2}$, $\lambda_p$
is the $p$-th order Lovelock coefficient and ${\cal{L}}_p$ is the Euler density of a $2p$--dimensional manifold.
In $d+1$ dimensions all ${\cal{L}}_p$ terms with $p\geq [{d\over 2}]$
are either total derivatives or vanish identically.

The Gauss-Bonnet action is the simplest example of a Lovelock action,
with only the 4--dimensional Euler density included. In the following we set $d=6$
\eqn\action{S=\int d^7 x  \sqrt{-g} {\cal{L}}=\int d^7 x  \sqrt{-g}\left(R+{30 \over L^2}+ {\lambda\over 1
2} L^2 {\cal{L}}_{(2)}\right)}
Note that in eq. \action\ we set $l_P=1$, included a cosmological constant term $\Lambda=-{30 \over L^2}$ and rescaled the
Lovelock parameter by $L^2$. The Gauss-Bonnet term ${\cal{L}}_{(2)}$ is
\eqn\gblang{{\cal{L}}_{(2)} =R_{MNPQ} R^{MNPQ}-4 R_{MN} R^{MN}+R^2 }
Equations of motion derived from \action\  can be expressed in the following way
\eqn\eqom{-{1\over 2}g_{MN}{\cal{L}}+R_{MN}+{1\over 6}\lambda {\cal{H}}^{(2)}_{MN} =0}
with ${\cal {H}}^{(2)}_{MN}$ defined as
\eqn\Hdef{{\cal{H}}^{(2)}_{MN}=R_{MLPQ} R_{N}^{\quad LPQ}-2 R_{MP}R_{N}^{\quad P}-2 R_{MPNQ}R^{PQ}+R R_{MN}}
Eq \eqom\ admits a solution of the form
\eqn\bhansatz{ds^2=-\tilde{a}^2 f(r)dt^2+{dr^2\over f(r)}+{r^2\over L^2} d\Sigma_{5,k}^2}
with $d\Sigma_{5,k}^2$ the metric of a $5$--dimensional manifold of constant curvature
equal to $20k$ \refs{\BoulwareDeser-\Cai}. Note that $\tilde{a}$ in eq. \bhansatz\ is an arbitrary constant which allows one
to fix the speed of light of the boundary theory to unity.
Given that we are interested in black hole solutions with flat horizon, we set $k=0$ in the following.
In this case the solution, known to be thermodynamically stable, reduces to\foot{Gauss-Bonnet gravity admits another AdS solution
with $\tilde{a}^2={1\over 2}\left[1-\sqrt{1-4\lambda}\right]$. This is however unstable and contains ghosts \BoulwareDeser.}
\eqn\bhsol{\eqalign{ds^2&=-\tilde{a}^2 f(r)dt^2+{dr^2\over f(r)}+{r^2\over L^2}\sum_{i=1}^{5}dx_i^2\cr
f(r)&={r^2\over L^2} X(r), \qquad
X(r)={1\over 2 \lambda}\left[1-\sqrt{1-4\lambda \left(1-{r_{+}^6\over r^6}\right)}\right]\cr
\tilde{a}^2&=\left[\lim_{r\ra\infty}{ {L^2\over r^2}f(r)}\right]^{-1}={1\over 2}\left[1+\sqrt{1-4\lambda}\right]
}}
The horizon is located at $r=r_{+}$ whereas the Hawking temperature of the black hole is
\eqn\temperature{T=\tilde{a} {3\over 2} {r_{+}\over L^2} }
In the limit ${r_+\over r}\ra0$ one recovers $AdS_7$ space from \bhsol.
The curvature scale of the AdS space is related to the cosmological constant via
\eqn\lads{   L_{AdS}=\tilde{a} L}

\newsec{Fluctuations}

\noindent
In this section we study fluctuations around the black hole solution \bhsol.
In particular, we consider small perturbations of the metric $h_{MN}$ in the scalar channel $\phi = h_{12}$.
In this case, the form of the perturbed metric is
\eqn\perds{ds^2=-\tilde{a}^2 f(r)dt^2+{dr^2\over f(r)}+{r^2\over L^2} \left[\sum_{i=1}^{5}dx_i^2+2 \phi(t,r,x_5)dx_1dx_2\right]}
Note that $\phi$ only depends on the $(t,r,x_5)$ directions of spacetime and its Fourier transform
can be written as
\eqn\Fourier{\phi(t,r, x_5)=\int {d\omega d q \over (2\pi)^2} \varphi(r) e^{-i\omega t+i q x_5} \qquad k=(\omega,0,0,0,0,q)}
The equations of motions for $\varphi$ can be found by substituting the ansatz \perds\ into \eqom\ and expanding to linear
order in the fluctuating field. The result is
\eqn\eqomfl{T_2 \varphi''(r)+T_1 \varphi'(r)+T_0 \varphi(r)=0}
with
\eqn\Tidef{\eqalign{T_0&=3 r \tom^2 \left[-2 r^3+\lambda L^2 r^2 f'(r)+2 r \lambda L^2 f(r)\right] +\cr
&+\tq^2 \tilde{a}^2 L^2 f(r)\left[-4\lambda L^2 r f'(r)+6 r^2 -\lambda L^2 r^2 f''(r)-2\lambda L^2 f(r)\right]\cr
T_1&=3 \tilde{a}^2 L^4 r f(r) \left[r^2 f'(r)\left(-2 r+\lambda L^2 f'(r)\right)
+6 \lambda L^2 f(r)^2  \right.\cr
&\left. \qquad\qquad\qquad  +r f(r) \left(8\lambda L^2 f'(r)-10 r^2 +r \lambda L^2 f''(r)\right) \right]\cr
T_2&=3 \tilde{a}^2 L^4 r f(r)^2\left[-2 r^3+\lambda L^2 r^2 f'(r)+2 r \lambda L^2 f(r)\right] }}
where primes indicate differentiation with respect to the variable $r$ and we defined $\tom=\omega L^2$
and $\tq=q L^2$.

It is convenient to place this equation in Schr\"{o}dinger form. To do this, we follow two steps:
We first define a new function $\Phi(r)$ through
\eqn\sone{\ln{\Phi}=\ln{\varphi}+{1\over 2}\int {T_1\over T_2}}
which brings eq. \eqomfl\ into the standard form
\eqn\eqomflsf{\Phi''(r)+W(r)\Phi(r)=0 \qquad W(r)={T_0\over T_2}-{1\over 2}\left[{T_1'T_2-T_2'T_1 \over T_2^2}\right]-{1\over 4}\left({T_1\over T_2}\right)^2 }
Eq. \Tidef\ then allows us to express $T_0(r)$ as $T_0(r)=b_{\omega}(r)\tom^2+b_{q}(r)\tq^2$
where $b_{\omega}(r),b_{q}(r)$ are functions of the radial coordinate $r$ alone.
This implies that $W(r)$ can be written as
\eqn\Whdef{W(r)={b_{\omega}\over T_2}\tom^2+{b_{q}\over T_2}\tq^2+h(r) \qquad h(r)=-{1\over 2}\left[{T_1'T_2-T_2'T_1 \over T_2^2}\right]-{1\over 4}\left({T_1\over T_2}\right)^2}
We then substitute $\Phi(r)$ with $\Psi(r)$ defined as $\Psi(r)=\left({b_{\omega}\over T_2}\right)^{{1\over 4}} \Phi(r)$
and subsequently make a coordinate transformation from $r$ to $y$ according to
\eqn\tvry{\p_r y(r)=\sqrt{{b_{\omega}\over T_2}}}
Eq. \eqom\ is finally expressed as
\eqn\eqomSchr{-\p_{y}^2\Psi+\left[\tq^2 c_g^2(y)+V_1(y)\right]\Psi=\tom^2 \Psi}
where
\eqn\cgsdef{c_g^2=-{ b_{q} \over b_{\omega} } \qquad\qquad
V(y)=-{T_2\over b_{\omega}}h(y)-\left({b_{\omega}\over T_2}\right)^{-{1\over 4}}
\p_y \left[ {b_{\omega}\over T_2} \p_y \left( {b_{\omega}\over T_2} \right)^{-{1\over 4}}\right] }

We are now ready to study the full graviton wave function \eqomSchr\ .
Note that $y(r)$ is a monotonically increasing function of $r$ with $y\ra 0$ at the boundary $r>>r_{+}$ and $y\ra -\infty$
at the horizon $r=r_{+}$. $V_1(y)$ blows up as $y^{-2}$ for $y\ra 0$.

Following \refs{\BriganteLMSYa,\BriganteLMSYb} we consider \eqomSchr\ in the limit $\tq\ra\infty$.
In this case, $\tq^2 c_g^2(y)$ provides the dominant contribution to the potential except for a small region $y> -{1\over \tq}$.
It is therefore reasonable to approximate the potential with $c_g^2(y)$ for all $y<0$ and replace it with an infinite wall
at $y=0$.
Consider now the behaviour of $c_g^2(y)$ near the boundary $y=0$. This is easier to
analyze in the original variable $r$. In particular,
\eqn\cgsex{c_g^2=1+C{r_{+}^6\over r^6}+\OO({1\over r^7})\qquad\qquad C=-{1-8\lambda +\sqrt{1-4\lambda } \over 2-8\lambda }}
Note that when $C$ is positive, $c_g^2(r)>1$
which implies (through WKB quantization) the existence of the
metastable states whose group velocity is greater than one
\BriganteLMSYb. Hence, for values of $\lambda$ such that $C>0$,
the boundary theory violates causality. That is, $C$ should remain negative for the dual
field theory to be consistent.
The values of the Gauss-Bonnet parameter $\lambda$ for which causality is preserved are determined from the
solutions of the inequality $C\leq 0$. To be precise,
\eqn\lambdabound{-{1-8\lambda +\sqrt{1-4\lambda} \over 2-8\lambda }\leq 0 \qquad\Rightarrow\qquad \lambda\leq {3\over 16}}
Note that although we have analyzed only the leading behaviour of $c_g^2$ close to the boundary
our results are exact since $c_g^2$ given by
\eqn\cgsfull{c_g^2=a^2 (1-4 \lambda) X(r) {(1-4 \lambda )+7\lambda {r^6_{+}\over r^6}\over \left[(1-4\lambda )+ \lambda {r^6_{+}\over r^6} \right] \left[(1-4\lambda )+4 \lambda {r^6_{+}\over r^6} \right] }}
is a monotonically increasing function of $r$ for all $\lambda \leq {3\over 16}$. $X(r)$ is defined in \bhsol.

This completes the discussion of this section. The gravity analysis imposes an upper bound \lambdabound\
on the Gauss-Bonnet parameter $\lambda$.
As we will see below, the AdS/CFT correspondence relates $\lambda$ to
the coefficients of the stress--energy tensor three point function of the boundary CFT.
Therefore, the bound on $\lambda$ can be translated into constraints on these coefficients.
In the next section we will consider how similar constraints arise in field theory.

\newsec{Energy flux one point functions and positivity of energy bounds}

\noindent
Consider the integrated energy flux per unit angle measured through a very large sphere
of radius $r$
\eqn\energydef{{\cal E}(\hat{n})=\lim_{r\rightarrow \infty}r^{d-2}\int dt \hat{n}^i T^0_ i(t, r\hat{n}^i)}
where $n^{i}$ denotes a unit vector in $R^{d-1}$, the space where the field theory lives. This unit vector
specifies the position on $S^{d-2}$ where energy measurements may take place. Integrating over
all angles yields the total energy flux at large distances.

An interesting object to consider \HofmanMaldacena\
is the energy flux one point function \SveshikovTkachov. It is defined as the expectation
value of the energy flux operator \energydef\ on states created by local operators ${\cal O}_q$
\eqn\vevenergydef{\vev {{\cal E}(\hat{n})}  ={{\bra 0}{\cal O}^{\dagger}_q {\cal{E}}(\hat{n}) {\cal O}_q {\ket 0}
\over {\bra 0} {\cal O}^{+}_q {\cal O}_q {\ket 0}  }}
The simplest case to examine is when the external states ${\cal O}_q {\ket 0}$ are produced by operators
with energy $q_0\equiv q$ and zero momentum ,\ie , $q^{\mu}=(q,0,0,0)$
\eqn\state{{\cal O}_q\equiv \int d^d x {\cal O}(x) e^{i q t} }
A state with generic four momentum $q^{\mu}$ can be obtained by performing a simple boost on
${\cal O}_q$ of \state.


Here we will be interested in six dimensional conformal field theories. In particular, we
will analyze the energy flux one point function on states produced by the stress--energy tensor operator
\eqn\ot{{\cal O}_q=\epsilon_{ij} T_{ij}(q)}
where $\epsilon_{ij}$ is a symmetric, traceless polarization tensor with indices purely in the spatial directions.
In complete analogy with \HofmanMaldacena\ $O(5)$ rotational symmetry allows us to express $\vev { {\cal E}(\hat{n})}$ as
\eqn\eqenergy{\vev{{\cal E}(\hat{n})}_{T_{ij}}={\vev {\epsilon^{*}_{ik}T_{ik}{\cal{E}}(\hat{n})\epsilon_{lj}T_{lj} } \over
\vev {\epsilon^{*}_{ik} T_{ik}\epsilon_{lj}T_{lj}} }=
{q_0\over \Omega_4}\left[1+t_2\left({\epsilon^{*}_{il}\epsilon_{lj}n_i n_j\over \epsilon^{*}_{ij}\epsilon_{ij}}-{1\over 5}\right)+
t_4 \left( {|\epsilon_{ij}n_i n_j|^2\over \epsilon^{*}_{ij}\epsilon_{ij} }-{2 \over 35}\right) \right]}
where $\Omega_4$ is the volume of the unit four-sphere $\Omega_4={8\pi^2\over 3}$. Hence, the energy flux one point
function is fixed by symmetry up to two coefficients, $t_2$ and $t_4$.

This result is in agreement with expectations from conformal invariance.
As explained in \OsbornPetkou, in any $d$--dimensional CFT the three point function of the stress--energy tensor can be
expressed in terms of three independent coefficients. They are denoted by $a,b,c$ in eqs. (3.15) to (3.21) of \OsbornPetkou.
On the other hand, the two point function of the stress--energy tensor depends on a unique parameter $C_T$
which is related to $a,b,c$ though
\eqn\CTdef{C_T=4 {2\pi^{{d\over 2}}\over \Gamma\left[{d\over2}\right]} {(d-2)(d+3)a-2b-(d+1)c \over d(d+2)} }
It is then convenient to ``change basis" and express the three point function in terms of $C_T$
and any other two linear combinations of $a,b,c$. Given that the energy flux one point function is actually
the ratio between a three- and a two- point function, it should  be completely determined up to
{\it two} independent parameters, \ie, the ratios of the two linear combinations of $a,b,c$ with $C_T$.

To obtain the numbers $(-{1\over 5},\,-{2\over 35})$ which appear in \eqenergy, we require that the integral
of the energy flux one point function over the four dimensional sphere yields the total energy $q=q_0$.
To see this one can use rotational invariance to set all components of the polarization tensor $\epsilon_{ij}$
to zero except for $\epsilon_{11}=-\epsilon_{22}$. Then
\eqn\enen{ {\epsilon^{*}_{il}\epsilon_{lj}n_i n_j\over \epsilon^{*}_{ij}\epsilon_{ij}}={1\over 2}\left(n_1^2+n_2^2\right)}
integrated over the four sphere and divided by its volume yields ${1\over 5}$.
The constant $-{2\over 35}$ in the last term of \eqenergy\ is obtained in an identical manner.

Much like in \HofmanMaldacena, positivity of the energy flux implies constraints on
the CFT parameters $t_2$ and $t_4$
\eqn\inequalities{\eqalign{&1-{1\over 5}t_2-{2\over 35} t_4\geq 0\cr
&\left(1-{1\over 5}t_2-{2\over 35}t_4\right)+{1\over 2}t_2 \geq 0 \cr
&\left(1-{1\over 5}t_2-{2\over 35}t_4\right)+{4\over 5}\left(t_2+t_4\right)\geq 0  }}
To obtain these inequalities
we use rotational symmetry to set $\hat{n}=\hat{x}_5$.
Then $\epsilon_{ij}$ can be separated into a tensor, vector and scalar
components with respect to rotations in $x^1\ldots x^4$.
The tensor component has $\epsilon_{5i}=\epsilon_{i5}=\epsilon_{55}=0$ and yields the first line in \inequalities.
The vector and scalar components give rise to
the second and third line respectively.
 Note that each of the
constraints \inequalities\ is saturated in a free field theory without antisymmetric tensor fields,
fermions or scalars respectively. This is similar to the situation
in four dimensions.

It is possible to calculate the energy one point function explicitly and thus derive the precise relations between
$t_2,t_4$ and the coefficients which appear in the stress-energy tensor three--point function $a,b,c$.
The computation is outlined in Appendix A. Here we present the results
\eqn\ttwotfour{\eqalign{t_2&={14\over 3} {220 a+ 54 b-39 c\over 36 a-2 b-7 c } \quad\qquad\Rightarrow\qquad
t_2=-140 {2 n_{a}- n_{f}\over 90 n_{a}+20 n_{f}+n_{s}}\cr
t_4&=-28 {65 a+ 24 b-14 c\over 36 a-2 b-7 c}\qquad\quad\Rightarrow\qquad t_4= {35\over 2}{6 n_{a}-8 n_f+n_s\over 90 n_a+20 n_f+n_s} }}
The expressions for free theories can be obtained with the help of \BastianelliFTa. First write $a,b,c$ of \OsbornPetkou\ in terms of
${\cal{A}},{\cal{B}},{\cal{C}}$ of \OsbornErdmenger\
\eqn\aAbBcC{a={{\cal A}\over 8},\qquad b={{\cal B}-2{\cal A} \over 8},\qquad c={{\cal C}\over 2}}
Then use \BastianelliFTa\ to relate ${\cal{A}},{\cal{B}},{\cal{C}}$ with $n_a,n_f,n_s$ ,\ie, the number of
free antisymmetric two--tensors, Dirac fermions and real scalars\foot{Note a factor of 2 missing from the last
term of the last line in eq.(3.22) of \BastianelliFTa\ for $d=6$.}.
\eqn\ABCfree{\eqalign{{\cal A}&=-{1\over \pi^3}\left[-{6^3\over 5^3}n_s+ {6^3\over 3}6 n_a\right] \cr
{\cal B}&=-{1\over \pi^3}\left[4 {6^3\over 5^3}n_s+{6^2\over 2}8 n_f+4 {6^3\over 3}6 n_a\right]\cr
 {\cal C}&=-{1\over \pi^3}\left[{4^2 6^2\over 4\cdot  5^3}n_s+ {6^2\over 4} 8 n_f+ 2 {6^3\over 3} 6 n_a\right]}}

Note  that for the supersymmetric $(2,0)$ multiplet with an anti-selfdual two form, five scalars and one Dirac
fermion, $t_4$ vanishes identically. In general, superconformal Ward identities result in an additional linear constraint on the
three point functions of the stress energy tensor  which reduces the number of independent parameters to two. 
(See \OsbornQU\ where the explicit form of the constraint is worked
out in four dimensional CFT.)
While the precise form of the constraint
is not known in six dimensions, it can be easily determined.
Recall that supersymmetry in six dimensions implies $6 n_a+n_s-8 n_f=0$
(this relation is satisfied by both scalar multiplet and (2,0) multiplet).
Using \ABCfree\ we can express this constraint in terms of ${\cal A}, {\cal B}, {\cal C}$ as
$17{\cal A}+24 {\cal B}-56 {\cal C}=0$.
This fixes the form of the constraint in six-dimensional superconformal theories.
Note that  supersymmetry in six dimensions implies $t_4=0$, just like in four dimensions \HofmanMaldacena.

With $t_4=0$ the inequalities in \inequalities\ reduce to
\eqn\susybounds{-{5\over 3}\leq t_2\leq 5}
thus constraining the domain of $t_2$ in any supersymmetric CFT.
Even when there is no supersymmetry, explicit bounds on $t_2,t_4$ can be derived from \inequalities.
In fact, in the space of $t_2$ and $t_4$ the solutions of \inequalities\ lie within a triangle
defined by the vertex points  $(-{28\over 9},{7\over 6}),(0,{35\over 2})$ and $(7,-7)$.

\newsec{Weyl anomaly and $t_2$ and $t_4$ from Gauss-Bonnet gravity}

\noindent
To compare the bounds \inequalities\ with the causality constraint \lambdabound\ we need
to compute $t_2$ and $t_4$.
More specifically, we need to determine the coefficients $a,b,c$
in a CFT hypothetically dual to GB gravity.
In principle, one can compute the three--point functions of the stress--energy
tensor directly by computing the scattering of three gravitons in the bulk of Anti de Sitter
space.
However this route is technically more challenging than the one
we take below.
Instead, we use the relation between $a,b,c$ and the coefficients
of the Weyl anomaly.
In a six-dimensional CFT the latter contains terms of three possible types
\eqn\wanomaly{  \AA_W= E_6 + \sum_i^3 b_i I_i +\nabla_i J^i}
where $E_6$ is the Euler density in six dimensions, $I_i, i=1,\ldots3$
are three independent conformal invariants composed out of the Weyl tensor
and its derivatives, and the last term is a total derivative of a covariant expression.
We will use $E_6$ and $I_i$ in the form quoted in \BastianelliFTb.

As explained in  \BastianelliFTb, the second term in $\AA_W$ (B-type anomaly)
comes from the effective action which contributes to the three-point
functions of the stress--energy tensor.
Hence, there is a linear relation between the coefficients $b_i$ in \wanomaly\
and the values of $a,b,c$.
We explain how to obtain this relation below.
To determine $b_i$ one needs to compute the Weyl anomaly in GB gravity.
The procedure for computing the Weyl anomaly of a CFT in a holographically
dual theory has been introduced in \HS.
Consider a $d$-dimensional CFT formulated on a space with Euclidean metric $g^{(0)}_{ij}$.
(We will be interested in the specific case $d=6$.)
Under a small Weyl transformation the metric changes as $\delta g^{(0)}_{ij}=2 \delta\sigma  g^{(0)}_{ij}$.
The CFT action is not invariant, but rather picks up an anomalous term,
\eqn\cftacweyl{  \delta W[ g^{(0)}_{ij}]=\int d^dx \sqrt{\det  g^{(0)} } \AA_W \delta\sigma  }
To compute the anomaly in GB gravity consider the following ansatz
for the metric \HS\
\eqn\adsanz{  ds^2=L_{AdS}^2 \left({1\over 4\rho^2} d\rho^2+{g_{ij}\over\rho} dx^i dx^j \right) }
where
\eqn\gijexp{ g_{ij}=g_{ij}^{(0)}+\rho g_{ij}^{(1)}+\rho^2 g_{ij}^{(2)}+\OO(\rho^3)  }
is an expansion  in powers of the radial coordinate $\rho$.
One can now solve the equations of motions of GB gravity order by order in the $\rho$
expansion and  determine $g_{ij}^{(i)}, i=1,\ldots$ in terms
of $g_{ij}^{(0)}$.
One should think of $g_{ij}^{(0)}$ as the metric which sources the stress--energy tensor
of the boundary CFT.
From the point of view of the AdS/CFT correspondence, specifying the source
completely determines the solution \adsanz.

To compute the anomaly, one needs to substitute the resulting
expansion \gijexp\ back into the lagrangian (more precisely, into $\sqrt{\det g}\, \LL$),  and extract the coefficient
of the $1/\rho$ term.
This is because when integrated over $\rho$ with the UV cutoff at $\rho=\epsilon$,
this term gives rise to a $\log \epsilon$ term in the six-dimensional
effective action.
This term is not removed by local counterterms and
gives rise to the anomaly \cftacweyl\ under Weyl transformation.
Clearly, the leading term in the expansion of the lagrangian
comes from $\sqrt{\det g}\sim 1/\rho^4$.
One may therefore be surprised that it is sufficient
to find $g_{ij}$ up to next-to-next-to leading order since an $\OO(\rho^3)$ term in \gijexp\
may naively contribute to the $\OO(\rho^{-1})$ term in $\sqrt{\det g}\, \LL$.
However one can check that  this term
[which is of the type ${g^{(0)}}^{ij} g^{(3)}_{ij}$] contains a multiplicative
factor which vanishes when the solution for the $AdS$ radius  \lads\ is substituted.
This has been observed previously in the context of gravity with $R^2$ terms
in four dimensions \NojiriMH.

In principle, the procedure outlined above can be performed analytically
to obtain the anomaly $\AA_W$ in the form \wanomaly.
However we opted to use the computer.
Consider the boundary metric of the form
\eqn\bdmetric{  g_{ij} dx^i dx^j = f(x^3,x^4) \left[ (dx^1)^2  +(dx^2)^2\right] +\sum_{i=3}^6 (dx^i)^2 }
One can check that for this metric $E_6=0$ and $I_i, i=1,\ldots3$ are linearly
independent combinations of terms which contain six derivatives of $f(x^3,x^4)$
distributed in various ways.
An example of such a term would be
\eqn\termexample{ [f^{(0,1)}(x^3,x^4)]^3 f^{(2,1)}(x^3,x^4)/[f(x^3,x^4)]^6   }
where $f^{(p,q)}\equiv (\p/\p x^3)^p (\p/\p x^4)^q f(x^3,x^4)$.
One can now use Mathematica to solve the equations of motion order by order in $\rho$.
The leading non-trivial term relates the value of the AdS radius with
the cosmological constant.
The next to leading term in the equations of motion determines $g^{(1)}$.
The non-vanishing components are $g^{(1)}_{11}=g^{(1)}_{22}$, $g^{(1)}_{33}$, $g^{(1)}_{44}$, $g^{(1)}_{34}=g^{(1)}_{43}$
$g^{(1)}_{55}=g^{(1)}_{66}$.
The explicit expressions can be obtained by using the following formula
\eqn\gijone{    g^{(1)}_{ij}= -{1\over4}\left(R_{ij}-{1\over10} R g^{(0)}_{ij}\right)  }
This is the result in Einstein-Hilbert gravity \HS.
It is not modified by the inclusion of the finite Gauss-Bonnet term.
The number of linearly independent algebraic equations at this order
is equal to the number of the nontrivial components of $g^{(1)}_{ij}$.
At the same order, there are more equations which contain derivatives
of $ g^{(1)}_{ij}$ with respect to $x^3,x^4$.
However upon substitution of \gijone\ these equations are identically
satisfied, which provides a good consistency check.
This story repeats itself at the next order as well.
However now $\lambda$ enters nontrivially into the solution
for $ g^{(2)}_{ij}$ which is somewhat cumbersome, so we will not
quote it here.

The next step involves substituting \adsanz\ together with the solution \bdmetric\
into the action \action\ and extracting the $1/\rho$ term in the integrand.
The resulting expression is too long to quote  here, but it must be
of the form $\int d^6 x \sqrt{\det g^{(0)}} \AA_W$, where  $\AA_W$ admits
the representation \wanomaly.
Both the anomaly $\AA_W$ and the invariants $I_i$ are long expressions involving
terms of the type \termexample.
Fortunately, the last (total derivative) term in the anomaly can also be
represented in a convenient  way.
In fact, any total derivative term must be of the form \BastianelliFTb
\eqn\totder{  \nabla_i J^i = \sum_{i=1}^7 c_i C_i     }
where $C_i$ are certain combinations of curvature invariants which
can be found in Appendix A of \BastianelliFTb, and $c_i$ are arbitrary
coefficients.
Now we compute $C_i$ for our choice of boundary metric \bdmetric\ and
demand that the coefficient in front of every term of the type \termexample\ in expression
\eqn\eqcoeff{ \AA_W-\sum_{i=1}^3 b_i I_i -\sum_{i=1}^7 c_i C_i =0 }
vanishes.
This uniquely fixes $b_i$ and $c_i$; the result is

\eqn\bi{\eqalign{  b_1 &= {208\over3}\left( -9(1+\sqrt{1-4\lambda})+2\lambda(31+22\sqrt{1-4\lambda}-52\lambda)\right)\cr
                   b_2 &= {70\over3} \left( -9(1+\sqrt{1-4\lambda})+2\lambda(35+26\sqrt{1-4\lambda}-68\lambda)\right)\cr
                   b_3 &= 70 (1+\sqrt{1-4\lambda}-2\lambda ) (1-4\lambda) \cr
 }}
and
\eqn\ci{\eqalign{  c_1 &= c_2=0\cr
                   c_3 &=  70 \left( 3 (1+ \sqrt{1-4\lambda})-2\lambda (11+8\sqrt{1-4\lambda})-20\lambda)\right)\cr
                   c_4 &= 84\left( -3 (1+\sqrt{1-4\lambda})+2 \lambda (11+8\sqrt{1-4\lambda}-20\lambda)\right)\cr
                   c_5 &= {140\over9}\left( -9(1+\sqrt{1-4\lambda})+\lambda(57+39\sqrt{1-4\lambda}-76\lambda)\right)\cr
                   c_6 &= 14 \left( -3 (1+\sqrt{1-4\lambda})+2\lambda(11+8\sqrt{1-4\lambda}-20\lambda)\right)\cr
                   c_7 &= {280\over3} \left(9(1+\sqrt{1-4\lambda})-2\lambda(30+21\sqrt{1-4\lambda}-40\lambda)\right)\cr
 }}
where we neglected an overall factor  common to all $b_i$'s and $c_i$'s
since only the ratios of $c_i$ will enter the expressions for $t_2$ and $t_4$ which we are after.
One can check that for $\lambda=0$ the expressions \bi\ and \ci\ reduce to
the numbers which appear in \BastianelliFTb.
In this case the values of $b_i$ are consistent with the
anomaly of a free (2,0) multiplet in six dimensions.
The coefficients in front of the total derivative terms are scheme-dependent
and therefore should not be compared.

The coefficients $b_i$ in \bi\ are related linearly to the parameters that determine
the two- and three--point functions of the stress--energy tensor, $\AA,\BB,\CC$.
To determine this relation we use the free field results for the Weyl
anomaly \BastianelliFTb:
\eqn\anomalyfree{  \AA_W{=} {-}\left({28\over3} n_s{+}{896\over3} n_f{+}{8008\over3}n_a\right)I_1+
           \left({5\over3}n_s{-}32 n_f{-}{2378\over3} n_a\right) I_2+\left(2 n_s{+}40 n_f{+}180 n_a\right) I_3  }
where $n_s,n_f,n_a$ are the same as those in eqs. \ttwotfour\ and \ABCfree.
In \anomalyfree\ we omitted the A-type anomaly term since
its coefficient is related to the four-point
function of the stress-energy tensor in $d=6$.
We have also omitted the total derivative term in \anomalyfree.
Using \anomalyfree\
together with the free field results for   $\AA,\BB,\CC$, \ABCfree,
we arrive at
\eqn\AABBCC{\eqalign{  \AA &= {864\over25}\left( 3(1+\sqrt{1-4\lambda})-\lambda(25+19\sqrt{1-4\lambda}-52\lambda)\right)\cr
                       \BB &= {72\over25}\left( 181(1+\sqrt{1-4\lambda})-\lambda(1275+913\sqrt{1-4\lambda}-2204\lambda)\right)\cr
                       \CC&=  {108\over25}\left( 59(1+\sqrt{1-4\lambda})-\lambda(425+307\sqrt{1-4\lambda}-756\lambda)\right)\cr
}}
Finally,  using \aAbBcC\ and \ttwotfour\ we determine the values of $a,b,c$, and  $t_2$ and $t_4$:
\eqn\ttwotfourres{   t_2= 5 \left({1\over\sqrt{1-4\lambda}}-1\right), \qquad t_4=0   }
As explained in the previous section,
vanishing $t_4$ is a necessary feature of any superconformal field theory.

We are now in position to substitute the values of $t_2$ and $t_4$ in \ttwotfourres\
into the constraints \inequalities\ and find out what the constraints are in terms
of $\lambda$.
The result is
\eqn\lambdares{  -{5\over16}\leq\lambda\leq{3\over16}   }
Note that the upper bound on $\lambda$ precisely coincides
with the upper bound \lambdabound\ obtained in Section 2
by demanding causality of the boundary theory.
The lower bound in \lambdares\ presumably can be obtained from
considering excitations with different polarization, just as it
has been in the four-dimensional setup \refs{\BuchelMyers-\Hofman}.

\newsec{Discussion}

\noindent In this paper we considered Gauss-Bonnet gravity in an $AdS_7$ background.
This theory has exact black hole solutions for values of the Gauss-Bonnet coupling $\lambda$
smaller than $1/4$.
We studied small fluctuations around these backgrounds and
showed that causality imposes an upper bound  \lambdabound\ on the
value of $\lambda$.
We expect a lower bound to follow from studying gravitons of different helicity.
To compare the causality bound with the positivity of energy bounds we computed $t_2$ and $t_4$
in a six-dimensional CFT in terms of the constants $a,b,c$ which
specify the three-point functions of the stress--energy tensor.
We then computed the holographic Weyl anomaly, and found that $t_4=0$.
The resulting bounds on $t_2$ are translated into bounds on $\lambda$ \lambdares.
Note that the upper bound in \lambdares\ is precisely the same as the causality bound \lambdabound.

We found that the results in six dimensions are very similar to those in
four dimensions.
The fact that $t_4=0$ is related to the truncation of the gravity action
at $\OO(R^2)$.
It would be interesting to include $R^3$ terms and see if the nonsupersymmetric bounds
can be addressed in this situation.
A natural generalization of the Gauss-Bonnet theory to $\OO(R^3)$ is 
third order Lovelock gravity.
Its equations of motion again contain only terms with at most second
derivatives acting on the metric.
There are two independent coefficients which multiply the Gauss-Bonnet
and the third order Lovelock term in the gravitational action.
Unfortunately, at least naively, the third order Lovelock term
does not contribute to the three-point functions of the stress-energy tensor,
and hence would not affect the value of $t_4$.
However the situation needs to be analyzed more carefully \WP.

One may view our result as an additional argument in favor of the
robustness of the correspondence between the positivity of energy and causality
conditions.
The understanding of this correspondence is at present somewhat incomplete
although it was argued in \Hofman\ that for Gauss-Bonnet gravity the equation for
a graviton propagating in a shock wave background receives contributions only from
the energy flux one-point function.
In fact, the form of eq. \lambdabound\ suggests that the leading small temperature
behaviour of the thermal two--point function might be determined by the three--point
function at zero temperature alone; and it is this leading term that determines causality
of the boundary theory.

Finally, let us consider the implications of our result on the viscosity to entropy ratio.
For a CFT hypothetically dual to $d+1$ dimensional Gauss-Bonnet gravity, this ratio has been computed in \BriganteLMSYa\
with the result\foot{Here $d\geq 4$ otherwise the Gauss-Bonnet term identically vanishes.}
\eqn\viscositytoentropy{{\eta\over s}={1\over 4\pi} \left(1-2 {d\over d-2} \lambda\right)
\quad \Longrightarrow_{d=6}\quad
{\eta\over s}={1\over 4\pi} \left(1-3\lambda\right)  }
Combining this with the bound on $\lambda$ leads to
\eqn\ratiobound{\left. {\eta\over s}\right|_{d=6} \geq {1\over 4\pi} {7\over 16}}
Note that the viscosity to entropy ratio is bounded from below by a number
smaller than that in four dimensions \BriganteLMSYb.
Hence, the lower bound on viscosity in Gauss-Bonnet gravity
depends on the dimensionality of the corresponding field theory.
More precisely, the bound decreases as one goes from $d=4$ to $d=6$.


This curious fact lead us to examine the restriction
causality imposes on $\eta/s$ for all $d+1$--dimensional Gauss-Bonnet
theories with $d\leq 10$.
The analysis is similar to the one done in Section 3.
We find that the viscosity
to entropy bound attains the smallest value when $d=8$:
\eqn\verdseven{\left. {\eta\over s} \right |_{d=8}\geq {1\over 4\pi}{219\over 529}}
It is interesting to note that the correspondence between causality and
positivity of energy also leads to an upper bound on viscosity in
the Gaus-Bonnet theories.
Perhaps further understanding of the bounds of the type \verdseven\ in generalized theories of gravity
may shed some light on the existence of a universal viscosity bound.

\bigskip
\bigskip

\noindent {\bf Acknowledgements:}
We thank F. Bastianelli, D. Kutasov, K. Papadodimas, L. Rastelli, K. Skenderis, M. Taylor, A. Tseytlin
and E. Verlinde for useful discussions and correspondence.
M.K. and A.P. thank the Aspen Center for Physics, the Centro de Ciencias de Benasque and the Simons
Center for Geometry and Physics where part of this work has been completed.
A.P. thanks the University of Amsterdam and the University of Chicago for hospitality.
This work was partly supported by NWO Spinoza grant and the FOM program {\it String Theory and Quantum Gravity}.

\appendix{A}{}

\noindent
To determine $t_2$ and $t_4$ with respect to the coefficients which appear in the three- point
function of the stress--energy tensor, an explicit computation of the energy flux is necessary.

Let us start with a careful consideration of eq. \eqenergy.
Without loss of generality we can use rotational symmetry to set the detector along the $\hat{x}_5$
direction. It is then convenient to define new coordinates $x^{\pm}=t\pm x_5$ and express the
energy flux measured at large distances as
\eqn\energyminus{{\cal E}=\lim_{x^{+}\rightarrow \infty} \left({x^{+}-x^{-}\over 2}\right)^4 \int dx^{-} T_{--}(x^{+},x^{-}) }
Since our main objective is to obtain $t_2, t_4$ it is sufficient to extract the
energy correlation function for two specific choices of the polarization tensor as long as they yield two
independent linear combinations of $t_2,t_4$. The simplest cases to consider are
$\epsilon_{ij}=0$ for all $i,j$ except for $\epsilon_{12}=\epsilon_{21}$ and $\epsilon_{ij}=0$ for all $i,j$ except for $\epsilon_{15}=\epsilon_{51}$. With these two choices, the following linear combinations of $t_2,t_4$ can be computed
\eqn\comba{1-{1\over 5}t_2-{2\over 35}t_4,\qquad 1+{3\over 10}t_2-{2\over 35}t_4}
In what follows we will analyze the former case in detail. The latter can be treated in an almost
identical manner.

In practice, we need to separately consider the numerator and denominator of eq. \eqenergy.
That is, we should compute the Fourier transform of the two- point function
\eqn\twopfourier{f_2(q_0)\equiv\int d^6x  e^{i q_0 t} \vev {T_{12}(x)T_{12}(0)}}
as well as the three- point function.
\eqn\threeptotal{f_3(q_0)\equiv\int d^6x e^{i q_0 t} \lim_{x_1^{+}\rightarrow \infty} \left({x_1^{+}-x_1^{-}\over 2}\right)^4 \int dx_1^{-}\vev {T_{12}(x)T_{--}(x_1)T_{12}(0) }}
As a warm up consider first the case of the two point function.
Its form is fixed by conformal invariance and according to \OsbornPetkou\ can be expressed as
\eqn\twopa{ \vev {T_{12}(x)T_{12}(0)}={C_T\over \left(x^2\right)^6} \left[ 1-{2\over x^2}\left(y_1^2+y_2^2\right)+{8\over \left(x^2\right)^2 } y_1^2 y_2 ^2\right]}
Here $x$ is a six vector parametrized as $x=\left(x^{+},x^{-},y_1,y_2,y_3,y_4\right)$ and $C_T$ satisfies eq. \CTdef\ with $d=6$.
To Fourier transform the above expression recall that the operators $T_{12}$ in \twopfourier\ are ordered as written.
This implies the $i\epsilon$ prescription $t\rightarrow t-i\epsilon$ which in light cone coordinates is replaced by
$x^{\pm}\rightarrow x^{\pm}-i\epsilon$. Integrating over $y_1,y_2$ using a spherical parametrization results in
\eqn\twopb{ f_2(q_0)={\pi^5\over 2\cdot 12\cdot 28} \left(36a-2b-7c\right) I_2 }
where we substituted $C_T$ in terms of the coefficients $a,b,c$ which determine the three point function
and denoted by $I_2$ the integral
\eqn\Itwodef{I_2= -{1\over 2} \int {dx^{+}\over \left(x^{+}-i\epsilon\right)^4}e^{i q_0 x^{+}\over 2}
\int {dx^{-}_2 \over \left(x^{-}-i\epsilon\right)^4}e^{i q_0 x^{-}\over 2}=
-{1\over 2} \left({1\over 3!}\right)^2(2 i\pi)^2 \left(i q_0\over 2\right)^6}

We are now ready to move on to the calculation of \threeptotal.
The starting point is once more the result of \OsbornPetkou\ where the form of the three- point function
of the stress--energy tensor is determined by conformal invariance up to three numbers $a,b,c$.
Adapting eq. (3.15) of \OsbornPetkou\ to the case of interest and taking the limit $\lim_{x_1^{+}\rightarrow \infty}$
yields
\eqn\threepa{\lim_{x_1^{+}\rightarrow \infty} \left({x_1^{+}-x_1^{-}\over 2}\right)^4 \vev {T_{12}(x) T_{--}(x_1)T_{12}(0)}=
{h(x)\over 64 \left(x^{-}_1-x^{-}+i\epsilon\right)^4 \left(x^{-}_1-i\epsilon\right)^4 \left(x^2\right)^6 }  }
with
\eqn\h{\eqalign{ h=\left(x^{-}\right)^2 \left \{ \left[\left(48 a+8 b-7 c\right) \left(y_1^4+y_2^4\right)+ c \left(y_3^2+y_4^2\right) +
\left(256 a+96 b-54 c\right)y_1^2 y_2^2 +\right.\right. &\cr
\left.\left.
+\left(48a+8b-6c\right)\left(y_3^2+y_4^2\right)\left(y_1^2+y_2^2\right)+2 c y_3^2 y_4^2\right] + \right. &\cr
\left.  +x^{-}x^{+} \left[-\left(48a+8b-6c\right) \left(y_1^2+y_2^2\right)-2 c \left(y_3^2+y_4^2\right)\right] +
c \left(x^{+}\right)^2\left(x^{-}\right)^2 \right\}& }}
Note that the $i\epsilon$ prescription here is such that the operator to the left of another acquires a more
negative imaginary part in the time direction. When integrating over $x_1^{-}$ we have the option of a contour
closing on either the upper or the lower $x_1^{-}$ plane thus including only one of the two poles in \threepa.
This results in
\eqn\threepa{\int dx_1^{-} \lim_{x_1^{+}\rightarrow \infty} \left({x_1^{+}-x_1^{-}\over 2}\right)^4  \vev {T_{12}(x_2) T_{--}(x)T_{12}(0)}=
{5i\pi\over 8} {h(x)\over \left(x^{-}-2 i\epsilon\right)^7 \left(x^2\right)^6 }}
Integrating now over the transverse coordinates $y_1,y_2$ yields
\eqn\threepc{f_3(q_0)={5i\pi^3 \over 8} {1\over 12}\left(28 a+6b-3c\right) I_3 }
where
\eqn\Ithreedef{ I_3=-{1\over 2}\int {dx^{+}_2\over \left(x^{+}_2\right)^2} e^{i q_0 x^{+}\over 2}
\int {dx^{-}_2 \over \left(x^{-}_2\right)^7} e^{i q_0 x^{-}\over 2}=-{1\over 2} {1\over 6!} \left(2i \pi\right)^2 \left({iq_0\over 2}\right)^7 }
Let us gather the results from eq.\twopb,$\,$\Itwodef,$\,$\threepc\ and \Ithreedef\ to form
the following ratio
\eqn\pref{{{8\pi^2\over 3}\over q_0}{f_3(q_0)\over f_2(q_0)}= -{7\over 3} {28 a+6b-3c\over 36a-2b-7c} }
As previously explained eq.\eqenergy\ combined with \pref\ leads to
\eqn\finalone{1-{1\over 5}t_2-{2\over 35} t_4=-{7\over 3} {28 a+6b-3c\over 36a-2b-7c}}
In a similar but slightly more complicated manner it is possible to compute the other linear combination of $t_2,t_4$ in \comba\
\eqn\finaltwo{1+{3\over 10}t_2-{2\over 35} t_4=28 {16a+4b-3c\over 36a-2b-7c } }
Solving then for $t_2,t_4$ reproduces eq. \ttwotfour.


\footatend\vfill\supereject\immediate\closeout\rfile\writestoppt
\baselineskip=14pt\centerline{{\bf References}}\bigskip{\frenchspacing%
\parindent=20pt\escapechar=` \input refs.tmp\vfill\eject}\nonfrenchspacing
\end